\documentclass{icrc29}
\usepackage{graphicx,amssymb,amsmath,times}
\usepackage{psfrag}
\begin{document}
\title[Simulations and parametrisation of radio emission from cosmic ray air showers]{Simulations and parametrisation of radio emission from cosmic ray air showers}
\author[T. Huege and H. Falcke] {T. Huege$^{a,b}$ and H. Falcke$^{b,c,d}$\\
(a) Institut\ f\"ur Kernphysik, Forschungszentrum Karlsruhe,
76021~Karlsruhe, Germany\\
(b) Max-Planck-Institut f\"ur Radioastronomie,
53010 Bonn, Germany \\
(c) Department of Astrophysics, Radboud University Nijmegen, 6525
ED Nijmegen, The Netherlands \\
(d) ASTRON, 7990 AA Dwingeloo, The Netherlands \\
        }
\presenter{Presenter: T. Huege (tim.huege@ik.fzk.de), \  
ger-huege-T-abs2-he13-poster}

\maketitle

\begin{abstract}

Cosmic ray air showers are known to emit pulsed radio emission at frequencies around a few tens of MHz. Accompanying the experimental efforts of the LOPES project, situated at the KASCADE-Grande site of the Forschungszentrum Karlsruhe, we have modeled the underlying emission mechanism in the scheme of coherent geosynchrotron radiation from electron-positron pairs deflected in the earth's magnetic field. As a follow-up to our earlier analytical calculations, we have developed a Monte Carlo simulation based on analytic parametrisations of air shower properties, including longitudinal and lateral particle distributions, particle energy and track-length distributions, and the longitudinal shower development as a whole. Here we present detailed simulation results. Important findings are the absence of significant asymmetries in the total field strength emission pattern in spite of the asymmetry introduced by the geomagnetic field, the polarisation characteristics of the geosynchrotron emission, allowing an unambiguous test of the geomagnetic emission mechanism, and the dependence of the radio emission on important shower parameters such as the shower zenith angle, the primary particle energy and the depth of the shower maximum. As a particularly useful result, these dependences have been summarized in a simple parametrisation formula, providing a solid basis for the interpretation of experimental data gathered with current and future experiments.

\end{abstract}

\section{Introduction}

Radio emission from extensive air showers (EAS) has been known to exist since its initial discovery in the 1960ies. Today's digital signal processing capabilities led to the idea of using digital radio interferometers such as LOFAR for its observation \cite{FalckeGorham}. These considerations sparked the LOPES project situated at the KASCADE-Grande site of the Forschungszentrum Karlsruhe in Germany, with the intention to develop a prototype for the measurement of radio emission from EAS based on LOFAR hardware. In the meantime, LOPES has successfully delivered proof-of-principle results confirming a geomagnetic origin of the emission \cite{LOPES}. In conjunction with the plans for an experimental design, a new approach for the interpretation of the emission mechanism was proposed \cite{FalckeGorham}. In this approach, the emission is interpreted as coherent geosynchrotron radiation arising from the deflection of the air shower cascade's secondary electron-positron pairs in the earth's magnetic field. Following this concept, we have carried out analytical calculations \cite{Analytics}, followed by detailed Monte Carlo simulations \cite{MonteCarlo} of the radio emission from cosmic ray air showers. The Monte Carlo simulations are based on an air shower model describing important shower properties such as the longitudinal and lateral particle distributions, the particle energy and track-length distributions and the overall longitudinal development of the air shower cascade with realistic, widely-used parametrisation formulas such as NKG-functions and Greisen-parametrisations, reaching a hitherto unprecedented level of detail in the simulations. Having verified the consistency of the analytical and Monte Carlo simulations, we now present a number of important results for the radio emission's dependence on specific air shower properties derived with this Monte Carlo model. A more complete discussion of these results can be found in \cite{Results}.

\section{Discussion}

\begin{figure}[ht]
   \psfrag{Eomegaew0muVpmpMHz}[c][t]{$|E_{\mathrm{EW}}(\vec{R},\omega)|$~[$\mu$V~m$^{-1}$~MHz$^{-1}$]}   
   \psfrag{nu0MHz}[c][t]{$\nu$~[MHz]}
   \centering
   \includegraphics[width=0.95\textwidth]{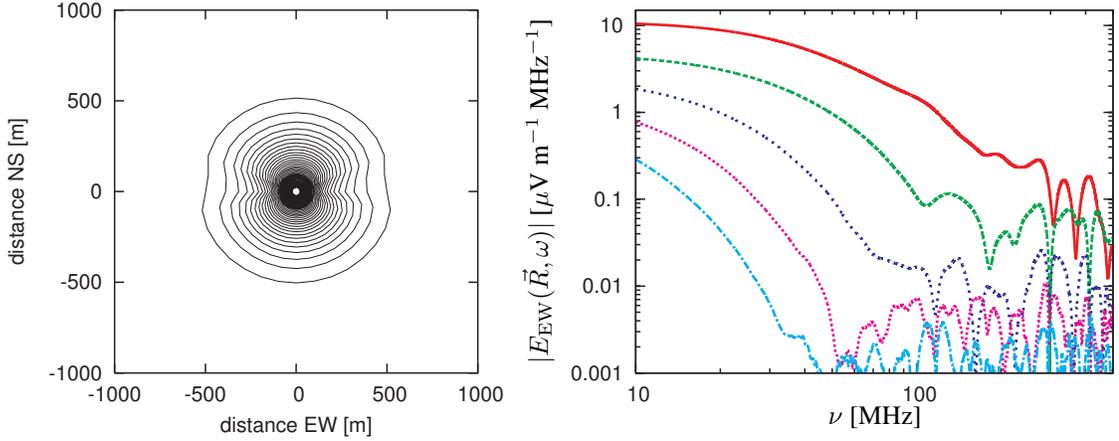}
   \caption[Radio emission from a vertical 10$^{17}$~eV air shower.]{
   \label{fig:vertical}
   10 MHz radio emission from a vertical $10^{17}\,$eV air shower. Left: Total field strength emission pattern. Right: Frequency spectra at (from top to bottom) 20$\,$m, 140$\,$m, 260$\,$m, 380$\,$m and 500$\,$m north of the shower centre. 
   }
   \end{figure}

To illustrate general radio emission properties, it is useful to first have a look at a very simple configuration, a vertical $10^{17}$~eV air shower. The magnetic field throughout this work is adopted as the one present in Central Europe, i.e., 70$^{\circ}$ inclined with a strength of 0.5 Gauss. Two very important characteristics of the radio emission from EAS are its spatial emission pattern and its frequency spectrum on the ground, shown in figure \ref{fig:vertical}. The total field strength emission pattern depicted in the left panel is remarkably circular. This is by no means self-evident, because the deflection of electrons and positrons is always directed in the east-west direction, making the emission of geosynchrotron radiation itself a highly asymmetrical process. The weakness of the asymmetry in the total field strength emission pattern is mainly due to the superior number of short particle tracks, which are much more numerous than long particle tracks that would lead to significant asymmetries.

The right panel of figure \ref{fig:vertical} shows the frequency spectra emitted by the vertical air shower as observed at various distances from the shower centre to the north. It is clear that the spectra cut off quickly to high frequencies, and measurements should thus concentrate on low frequencies if allowed by the noise background. The spectral cutoff is caused by the loss of coherence once the wavelength of the emission becomes comparable to or smaller than the thickness of the air shower disk. The effect becomes more dramatic when one goes to higher distances due to projection effects enlarging the apparent thickness of the disk. At high frequencies, especially far away from the shower centre, one enters the incoherent regime and the Monte Carlo simulations show an unphysical seeming, rapidly alternating series of interference minima and maxima. These are an artifact of the total homogeneity of the particle distributions in the simulated showers. To make statements about the emission levels in the incoherent regime (which could well be of measurable strength), one therefore needs a better air shower model, as is currently being implemented in the Monte Carlo code \cite{Talk} using particle distributions derived from CORSIKA \cite{CORSIKA} simulations.

\begin{figure}[ht]
   \psfrag{Eomegaew0muVpmpMHz}[c][t]{$|E_{\mathrm{EW}}(\vec{R},\omega)|$~[$\mu$V~m$^{-1}$~MHz$^{-1}$]}   
   \psfrag{nu0MHz}[c][t]{$\nu$~[MHz]}
   \centering
   \includegraphics[width=0.95\textwidth]{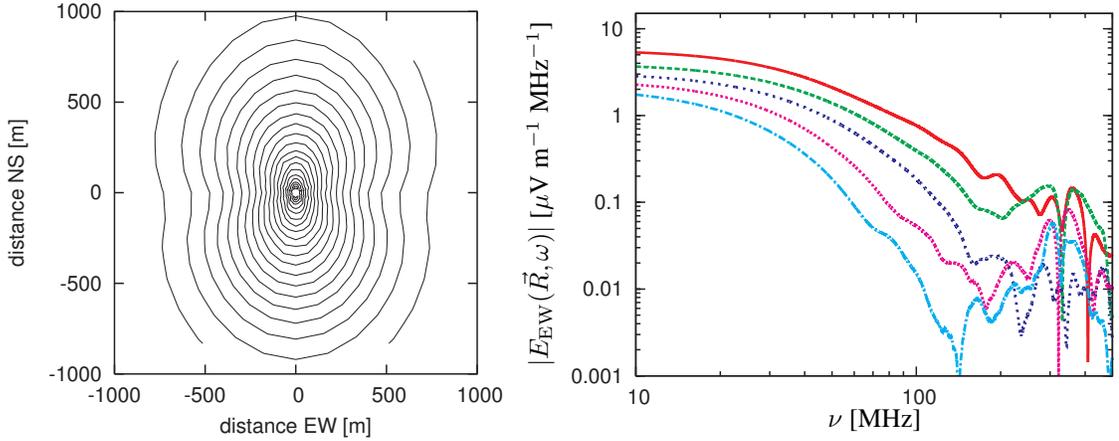}
   \caption[Radio emission from an inclined 10$^{17}$~eV air shower.]{
   \label{fig:inclined}
   10 MHz radio emission from a 45$^{\circ}$ inclined $10^{17}\,$eV air shower. Left: Total field strength emission pattern. Right: Frequency spectra at (from top to bottom) 20$\,$m, 140$\,$m, 260$\,$m, 380$\,$m and 500$\,$m north of the shower centre.
   }
   \end{figure}

The picture changes when one looks at an inclined air shower. Figure \ref{fig:inclined} shows again the total field strength emission pattern and the frequency spectra, but now for the radio emission from a $10^{17}$~eV air shower with 45$^{\circ}$ zenith angle coming from the south. Comparing the emission pattern with that of figure \ref{fig:vertical}, one immediately notices a strong projection effect, elongating the pattern along the shower axis. Apart from this intuitive projection effect, the pattern also generally gets broader, even in the direction perpendicular to the shower axis. This is a consequence of the shower maximum's much larger spatial distance to the observer for an inclined shower as compared with a vertical shower. (The same effect arises when varying the depth of shower maximum for a fixed air shower geometry \cite{Results}.) Their much broader emission pattern thus makes inclined showers a specifically suitable target for observations of the radio emission from EAS, see also \cite{Jelena} and \cite{Franzosen}. Comparing the frequency spectra of the inclined shower (right panel of figure \ref{fig:inclined}) with those of the vertical shower (right panel of figure \ref{fig:vertical}) demonstrates a similar effect. For the inclined air shower, the emission stays coherent up to much higher distances from the shower centre, again allowing much easier detection of inclined showers compared with vertical showers.

\begin{figure}[!ht]
   \psfrag{Eomegaew0muVpmpMHz}[c][t]{$|E_{\mathrm{EW}}(\vec{R},2\pi\nu)|$~[$\mu$V~m$^{-1}$~MHz$^{-1}$]}   
   \centering
   \includegraphics[width=0.93\textwidth]{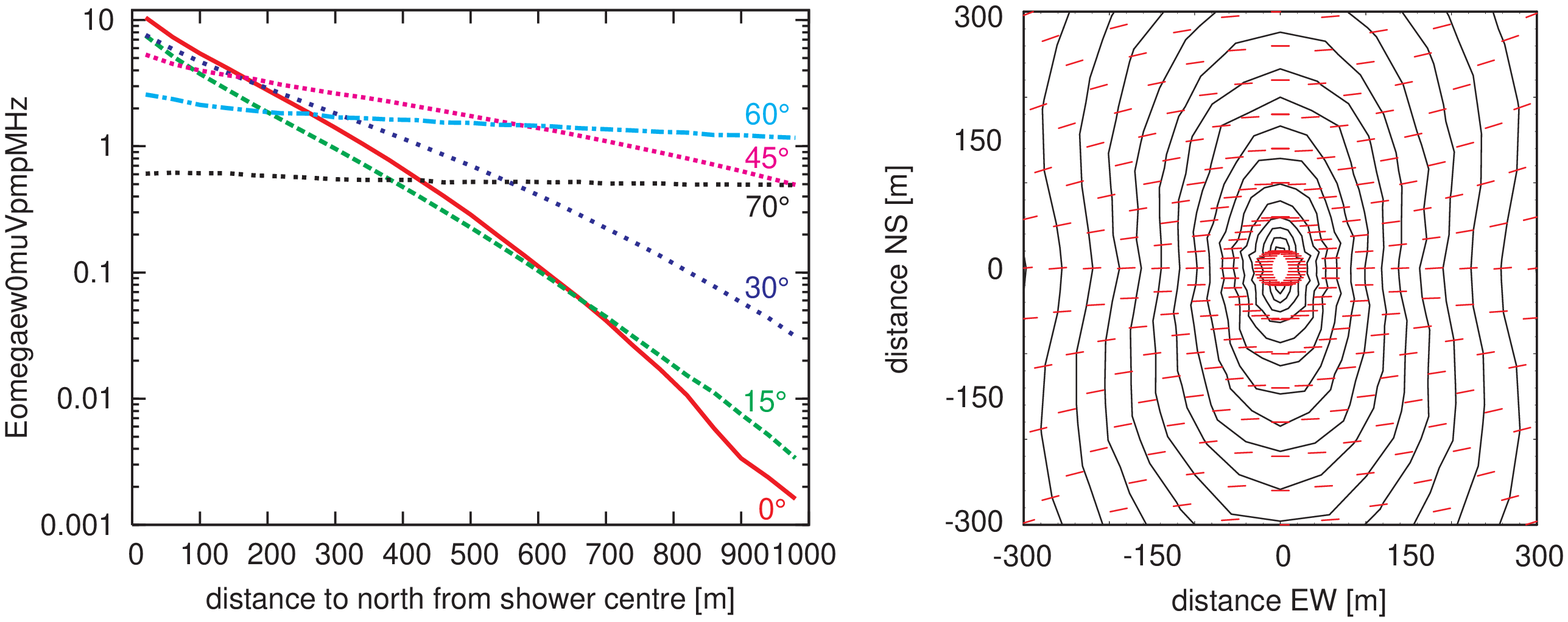}
   \caption[Inclination and Polarisation]{
   \label{fig:inclinationandpol}
   Left: Radial dependence of the 10~MHz electric field strength emitted by 10$^{17}$~eV air showers with various zenith angles (coming from the south). Right: Zoomed in total field strength emission pattern of a 10$^{17}$~eV air shower with 45$^{\circ}$ zenith angle overplotted with indicators denoting the ratio of east-west to north-south linear polarisation.
   }
   \end{figure}

A more quantitative view of the air shower inclination's effects on the emission pattern is given in the left panel of figure \ref{fig:inclinationandpol}. It shows the 10~MHz electric field strength as a function of radial distance to the north of the shower centre (i.e., cuts through the emission pattern along the shower axis). While the effect is only minimal up to $\sim$~15--20$^{\circ}$, the pattern becomes much more elongated and broader at zenith angles greater than $\sim$~30$^{\circ}$.

Another very important result of the simulations are the polarisation characteristics of the emission. While the total field strength emission pattern only exhibits a very subtle dependence on the magnetic field direction, separate measurements of the north-south and east-west polarisation components should show a significant dependence on the shower azimuth angle relative to the magnetic field. The simulations predict that the emission is predominantly linearly polarised. The right panel of figure \ref{fig:inclinationandpol} shows a blow-up of the total field strength emission pattern of the 45$^{\circ}$ inclined air shower. Overlaid are indicators denoting the ratio of east-west to north-south polarisation. (Horizontal indicators denote pure east-west polarisation, vertical ones pure north-south polarisation.) The figure demonstrates that close to the shower centre, the emission is predominantly polarised in the direction perpendicular to the magnetic field and shower axis, which was also a result of the analytic calculations \cite{Analytics} and historical works investigating a geomagnetic emission mechanism. For other possible radio emission mechanisms such as Askaryan-type \v{C}erenkov radiation which plays a dominant role in dense media, the expected polarisation characteristics are different. Linear polarisation measurements, as are possible with LOPES, can thus verify the geomagnetic origin of the emission in a very direct way.

The simulations also confirm the expected linear scaling of the electric field amplitudes with the primary particle energy, as long as one observes in the coherent regime, where the emission from all individual particles adds up coherently and thus scales linearly with particle number. Some deviations from the linear scaling arise through the additional effect that higher energy showers on average penetrate deeper into the atmosphere than lower energy ones. The overall effect is demonstrated again in \cite{Results}.

To give a summary of the aforementioned and additional dependences of the radio emission on the underlying air shower parameters, we have derived a simple, analytical parametrisation formula reproducing the Monte Carlo results with good accuracy \cite{Results}. This parametrisation has been implemented as a web-based online-calculator that can be a useful tool for estimates of the radio emission.


\section{Conclusions}

We have established important characteristics of the radio emission from EAS and their dependence on the properties of the underlying air showers. Among the most important results we demonstrate that inclined air showers pose a particularly interesting target for radio observations due to their larger emission pattern and better coherence up to high frequencies. We demonstrate that linear polarisation measurements can be used to directly verify the geomagnetic origin of the emission. A simple parametrisation formula summarises our results in a particularly useful way. Our Monte Carlo code is currently being equipped with a CORSIKA-based air shower model, allowing further improvements in the modelling quality.


\end{document}